\documentclass[letterpaper,pteplogo]{ptephy}

\usepackage[caption=false]{subfig}

\setcitestyle{numbers}

\newcommand{\ve}{\varepsilon}
\newcommand{\rme}{\mathrm{e}}
\newcommand{\rmi}{\mathrm{i}}
\newcommand{\rmd}{\mathrm{d}}

\newcommand{\calN}{\mathcal{N}}
\newcommand{\calD}{\mathcal{D}}
\newcommand{\calJ}{\mathcal{J}}
\newcommand{\calO}{\mathcal{O}}
\newcommand{\hdelta}{\hat{\delta}}

\renewcommand{\Re}{{\rm Re}\,}
\renewcommand{\Im}{{\rm Im}\,}

\begin{document}

\title{Hamilton dynamics for the Lefschetz thimble integration akin to
 the complex Langevin method}

\author{
\name{\fname{Kenji} \surname{Fukushima}}{1},
\name{\fname{Yuya} \surname{Tanizaki}}{1,2}
}

\address{
\affil{1}{Department of Physics, The University of Tokyo,
          7-3-1 Hongo, Bunkyo-ku, Tokyo 113-0033, Japan}
\affil{2}{Theoretical Research Division, Nishina Center, RIKEN
          Wako 351-0198, Japan}}

\begin{abstract}
  The Lefschetz thimble method, i.e., the integration along the
  steepest descent cycles, is an idea to evade the sign problem by
  complexifying the theory.  We discuss that such steepest descent
  cycles can be identified as ground-state wave-functions of a
  supersymmetric Hamilton dynamics, which is described with a
  framework akin to the complex Langevin method.  We numerically
  construct the wave-functions on a grid using a toy model and confirm
  their well-localized behavior.
\end{abstract}

\subjectindex{A22, A24, B15}

\maketitle

\paragraph{1.\ Introduction.}

The first-principle approach to solve the theory is an ultimate goal
of theoretical investigations.  The quantum Monte Carlo simulation for
the functional integral has been a powerful \textit{ab initio}
technique to reveal non-perturbative features of various physical
systems.  Successful applications include the lattice-QCD (Quantum
Chromodynamics) simulation, the lattice Hubbard model, the path
integral representation of spin systems, etc.

The Monte-Carlo algorithm is based on importance sampling, so it is
demanded that the integrand should be positive semi-definite, i.e.,
$\rme^S\ge0$ where $S$ is the classical action.  This positivity
condition is often violated in systems of our interest and then we can
no longer rely on Monte Carlo simulations~\cite{Loh:1990zz,Muroya:2003qs}.  In QCD a finite baryon or
quark density introduces a mixture of Hermitian and anti-Hermitian
terms in the action, and then $\rme^{S}$ acquires a complex phase.  In
repulsive Hubbard model away from half-filling or generally in
fermionic systems with spin imbalance, the sign of the integrand may
fluctuate.  It is also a notorious problem of the complex phase that
appears from the Berry curvature in the path integral representation
of spin systems and this phase cannot be removed for frustrated
situations such as the XY model on the Kagom\'{e} lattice.

Moreover, to approach real-time quantum phenomena, $\rme^S$ is an
oscillating function by definition and the sign problem is
unavoidable.  Although the Monte Carlo simulation is useful to compute
physical observables in equilibrium in the imaginary-time formalism,
the analytical continuation is necessary to access the real-time
information.  In general, however, the analytical continuation is a
quite pricey procedure, and some additional information on the system
such as the pole and the branch-cut structures would be necessary. 

Many ideas have been proposed so far to overcome the sign problem and,
unfortunately, applicability of each method has been severely
limited.  Recently new techniques to complexify the theory are
attracting more and more theoretical interest, which includes the path
integral on Lefschetz
thimbles~\cite{pham1983vanishing, Witten:2010cx, Witten:2010zr,%
  Cristoforetti:2012su,Cristoforetti:2013wha, Fujii:2013sra, Cristoforetti:2014gsa,%
  Tanizaki:2014xba,Tanizaki:2014tua, Kanazawa:2014qma,%
  Tanizaki:2015pua, DiRenzo:2015foa}
and the complex Langevin approach~\cite{Damgaard:1987rr,Aarts:2010aq,%
  Makino:2015ooa,Nishimura:2015pba,Berges:2005yt,Berges:2006xc,%
  Anzaki:2014hba}.
Except for several formal arguments, theoretical foundations for
the complex Langevin method are not fully established, and not much
is known about its reliability~\cite{Aarts:2010aq,Makino:2015ooa,Nishimura:2015pba}.
In the context of real-time quantum systems, the numerical simulation
works for some initial density
matrices~\cite{Berges:2005yt,Berges:2006xc}; however, it is recently
reported in Ref.~\cite{Anzaki:2014hba} that the real-time anharmonic
oscillator at zero temperature converges to a wrong answer with
unphysical width.

In contrast, the Lefschetz-thimble method has a solid mathematical
foundation at least for finitely multiple
integrals~\cite{pham1983vanishing,Witten:2010cx,Witten:2010zr};
however, its practical applicability is still in the developing
stage.  This method decomposes the original integration cycle into
several steepest descent ones, called Lefschetz thimbles, using
complexified field variables.
Picking up a single Lefschetz thimble, one can employ importance
sampling~\cite{Cristoforetti:2012su,Cristoforetti:2013wha,Fujii:2013sra}
and one can evade the sign problem since the oscillatory factor in
${\rme^S}$ totally disappears on each Lefschetz thimble.  On the other
hand, zero-dimensional model studies have exemplified the importance
of structures of multiple Lefschetz
thimbles~\cite{Tanizaki:2014xba,Tanizaki:2014tua,Kanazawa:2014qma,%
  Tanizaki:2015pua}.
For further applications it is needed to deepen our understanding on
more aspects of the Lefschetz thimbles.

The purpose of this Letter is to shed new light on the Lefschetz
thimble method in a form of the Hamilton dynamics, which was first
elucidated in Ref.~\cite{Witten:2010zr}.  In this reformulation, the
Lefschetz thimbles can be identified as ground-state
wave-functions of a supersymmetric topological quantum system.
After reviewing this modified Lefschetz thimble method, for a
  quartic potential problem at zero dimension, we solve the Hamilton
dynamics concretely to find the corresponding wave-functions.  Based
on our analytical and numerical observations, we discuss
advantages of this reformulation for the numerical computation.

\paragraph{2.\ Lefschetz thimble and SUSY quantum mechanics.}

Let us consider an $N$-dimensional real integral as a ``quantum field
theory'' defined by a (complex) classical action $S(x)$. 
In this theory our goal is to compute an expectation value
of an ``observable'' $\calO(x)$ defined by
\begin{equation}
  \langle \calO\rangle = \calN\int_{-\infty}^\infty \rmd^N x\;
  \rme^{S(x)}\, \calO(x) \;,
\label{eq:original}
\end{equation}
where $x=(x^{(1)},x^{(2)},\dots,x^{(N)})\in\mathbb{R}^N$ and the
normalization $\calN$ is chosen such that $\langle 1\rangle = 1$.  The
starting point in our discussion is to reformulate this theory in an
equivalent and more treatable way using a complexified representation:
\begin{equation}
  \langle \calO\rangle = \int \rmd^N z\,\rmd^N\bar{z}\;
  P(z,\bar{z})\, \calO(z) \;.
\label{eq:complexified}
\end{equation}
Here $z^{(i)}=x_1^{(i)}+\rmi x_2^{(i)}$ and
$\bar{z}^{(i)}=x_1^{(i)}-\rmi x_2^{(i)}$ with
$x_1^{(i)},x_2^{(i)}\in\mathbb{R}$ and $\int\rmd z\rmd \bar{z}$
represents the integration over the whole complex plane; i.e.\
$\int_{-\infty}^\infty \rmd x_1 \int_{-\infty}^\infty \rmd x_2$.  The
choice of the generalized weight function $P(z,\bar{z})$ may not be
unique.  Indeed, a trivial example is
$P(z,\bar{z})=\calN\rme^{S(z)}\prod_i\delta(z^{(i)}-\bar{z}^{(i)})$.
At the cost of complexifying the variables, nevertheless, it is often
the case that $P(z,\bar{z})$ could be endowed with more desirable
properties for analytical and numerical computation than the original
$\rme^{S(x)}$.

A clear criterion to simplify the integral is to find $P(z,\bar{z})$
such that the phase oscillation can be as much suppressed along
integration paths as possible, while in the complex Langevin
  method $P(z,\bar{z})$ is optimized to become a real probability.
To suppress the phase oscillation, let us pick up a saddle point
$z_\sigma$ satisfying $S'(z_\sigma)=0$.  The steepest descent
cycle or the Lefschetz thimble $\calJ_\sigma$ of the
saddle point $z_\sigma$ is defined with a fictitious time $t$ as
\begin{equation}
 \calJ_\sigma = \biggl\{ z(0)=x_1(0)+\rmi x_2(0) \;\biggr|\;
  \frac{\rmd x_j^{(i)}(t)}{\rmd t}=-\frac{\partial \Re S}{\partial x_j^{(i)}},\;
  \lim_{t\to-\infty}(x_1 (t)+\rmi x_2(t)) = z_\sigma \biggr\}\;.
  \label{eq:thimble}
\end{equation}
This is a multi-dimensional generalization of the steepest descent
path in complex analysis, which we will refer to as the
\textit{downward} path.  The original integration path on the real
axis {in} Eq.~\eqref{eq:original} can be deformed as a sum of
contributions on $\calJ_\sigma$ weighted with an integer $m_\sigma$;
i.e.,
$\int_{\mathbb{R^N}}\rmd^N x =\sum_{\sigma}m_\sigma \int_{\calJ_\sigma}\rmd^N z$.
In mathematics it is established how to determine $m_\sigma$ from the
intersection pattern between the steepest ascent (\textit{upward})
path from $z_\sigma$ and the original integration
path~\cite{pham1983vanishing,Witten:2010cx,Witten:2010zr}.  It is
important to note that $\Im S$ is a constant on each Lefschetz thimble
for the application to the sign
problem~\cite{Cristoforetti:2012su,Fujii:2013sra}.

In the following, let us restrict ourselves to $N=1$ for simplicity, because the generalization is straightforward. 
So far, the Lefschetz thimble is constructed as a line, and let us
find a two-dimensional smooth distribution $P(z,\bar{z})$ according to
Ref.~\cite{Witten:2010zr}.  
For that purpose, we define the ``delta-functional one-form'' $\delta(\mathcal{J}_{\sigma})$ supported on the Lefschetz thimble so that 
\begin{equation}
\int_{\mathcal{J}_{\sigma}}\mathcal{O}(z)\mathrm{e}^{S(z)}\rmd z=\int_{\mathbb{C}}\delta(\mathcal{J}_{\sigma})\wedge \mathcal{O}(z) \mathrm{e}^{S(z)}\rmd z. 
\label{Eq:IntegrationOnLefschetzThimble}
\end{equation}
For instance, $\delta(\mathbb{R})=\delta(y)\rmd y$. 
Such delta-functional forms $\delta(\mathcal{J}_{\sigma})$ (on a K\"ahler manifold) have a path-integral expression from the supersymmetric quantum mechanics \cite{Frenkel:2007ux, Frenkel:2006fy, Frenkel:2008vz} (see also Secs.~2.8 and~4 of Ref.~\cite{Witten:2010zr} for more details in this context). 
Integration (\ref{Eq:IntegrationOnLefschetzThimble}) can be represented as
\begin{eqnarray}
  \langle \calO\rangle &=& \calN\int \calD[x,p,\pi,\psi]\;
  \exp\biggl[\rmi
  \int_{-\infty}^0 \rmd t\, p_i \biggl( \frac{\rmd x_i}{\rmd t}
  + \frac{\partial \Re S}{\partial x_i} \biggr)\biggr]\nonumber\\
&&\times
  \exp\biggl[ -\int_{-\infty}^0 \rmd t\,
  \pi_i\biggl( \frac{\rmd}{\rmd t}\delta_{ij}+\frac{\partial^2 \Re S}
  {\partial x_i\partial x_j}\biggr)\psi_j\biggr]
  \;  \calO(z(0))\, \rme^{S(z(0))}\,(\psi_1\!+\!\rmi\psi_2)(0)\;.
\label{eq:path_integral_expression_thimble}
\end{eqnarray}
Here $x,p$ are bosonic fields and $\pi, \psi$ are fermonic ghost fields, and $z(t)\to z_{\sigma}$ as $t\to -\infty$. 
We should note that an integration in terms of $z$ is promoted to the path integral on $z(t)$ for $t\le 0$, 
while the observable and the weight $\calO(z(0))\exp S(z(0))$ are functions of $z(0)$ only.  
Let us outline how these two expressions (\ref{Eq:IntegrationOnLefschetzThimble}) and (\ref{eq:path_integral_expression_thimble}) are equivalent~\cite{Witten:2010zr, Frenkel:2007ux, Frenkel:2006fy, Frenkel:2008vz}. We first integrate out $p(t)$ to get the Dirac delta function,
\begin{equation}
\int\calD p\, \exp\biggl[\rmi
  \int_{-\infty}^0 \rmd t\, p_i \biggl( \frac{\rmd x_i}{\rmd t}
  + \frac{\partial \Re S}{\partial x_i} \biggr)\biggr]
= \delta\biggl(\frac{\rmd x_i}{\rmd t} + \frac{\partial \Re S}
  {\partial x_i}\biggr)\;.
\label{eq:thimble_delta_functional}
\end{equation}
This delta function constrains the path integral on $x(t)$ to a gradient-flow line defining Lefschetz thimbles. Since $z(-\infty)\to z_{\sigma}$, this path integral for $t<0$ gives a delta-functional support on $\mathcal{J}_{\sigma}$. 
However, the delta function produces an unwanted determinant factor. 
As is well-known, the path integral on ghost fields $\pi(t),\psi(t)$ for $t<0$ can eliminate that factor as 
\begin{equation}
\int\calD \pi\,\calD\psi\, \exp\biggl[ -\int_{-\infty}^0 \rmd t\,
  \pi_i\biggl( \frac{\rmd}{\rmd t}\delta_{ij}+\frac{\partial^2 \Re S}
  {\partial x_i\partial x_j}\biggr)\psi_j\biggr]
=  {\rm Det}\biggl( \frac{\rmd}{\rmd t}\delta_{ij}
  +\frac{\partial^2 \Re S}{\partial x_i\partial x_j}\biggr)
\;.
\label{eq:thimble_functional_determinant}
\end{equation}
Now, we obtain an integration over surface variables $x(0),\, \psi(0)$, and denote them by $x,\,\psi$. 
Locally, the Lefschetz thimble $\mathcal{J}_{\sigma}$ can be expressed as zeros of a certain function $f$, then we can find that the path integral (\ref{eq:path_integral_expression_thimble}) eventually gives
\begin{equation}
\int \rmd^2 x \rmd^2 \psi\, \delta(f)\, {\partial f\over \partial x_i}\psi_i \wedge \calO(z)\, \rme^{S(z)}\,(\psi_1\!+\!\rmi\psi_2)=\int \delta(f(x))\rmd f(x)\wedge \calO(z)\, \rme^{S(z)}\,\rmd z, 
\end{equation}
which is nothing but the local expression of the original integration (\ref{Eq:IntegrationOnLefschetzThimble}). 
Going back to (\ref{eq:path_integral_expression_thimble}), this shows that the so-called residual sign problem comes from the fermionic surface term $\psi_1(0)+\rmi \psi_2(0)$ because one can identify
$\psi_i(0)=\rmd x_i$ as above. 

Importantly, with these added fields, $p_i, \pi_i, \psi_i$, the action
is BRST exact under a transformation; $\hdelta x_i=\psi_i$,
$\hdelta\psi_i=0$, $\hdelta\pi_i=-\rmi p_i$, $\hdelta p_i=0$.  By
definition the nilpotency $\hdelta^2=0$ is obvious.  Thanks to the
boundary fermionic operator in
\eqref{eq:path_integral_expression_thimble}, the surface term is
BRST-closed so long as the observables are holomorphic.  This
  makes a sharp contrast to the complex Langevin method that could
  also acquire the BRST symmetry but it is violated by the surface
  term.
Because of the BRST symmetry we can add any BRST exact terms without
changing the original integral, and it is useful to insert
$\frac{\ve_i}{2}\int \rmd t\,p_i^2$.  In summary, the
effective Lagrangian that describes the fictitious time evolution is
given by the following topological theory:
\begin{equation}
\begin{split}
  L_{\rm eff} &= -\frac{\ve_i}{2}p_i^2 + \rmi p_i
  \biggl(\frac{\rmd x_i}{\rmd t}+\frac{\partial \Re S}{\partial x_i}
  \biggr) + \pi_i\biggl(\frac{\rmd}{\rmd t}\delta_{ij}
  + \frac{\partial^2 \Re S}{\partial x_i \partial x_j}\biggr)\psi_j\\
 &= -\hdelta\, \left\{\pi_i\biggl(\rmi\frac{\ve_i}{2}p_i
  +\frac{\rmd x_i}{\rmd t}+\frac{\partial\Re S}{\partial x_i}\biggr)\right\}\;,
\end{split}
\label{eq:Leff}
\end{equation}
which is nothing but a Legendre transform of an effective Hamiltonian:
\begin{equation}
  H_{\rm eff} = \sum_{i}\biggl[\frac{\ve_i}{2} \hat{p}_i^2
  -\frac{\rmi}{2} \biggl( \frac{\partial \Re S}{\partial x_i}\hat{p}_i
  + \hat{p}_i \frac{\partial \Re S}{\partial x_i}\biggr)\biggr]
  - \sum_{i,j} \frac{1}{2}\frac{\partial^2 \Re S}
  {\partial x_i \partial x_j} \bigl[ \hat{\pi}_i, \hat{\psi}_j\bigr]
\label{eq:Heff}
\end{equation}
with $[x_i,\hat{p}_j]=\rmi \delta_{ij}$ and
$\{\hat{\pi}_i,\hat{\psi}_j\}=\delta_{ij}$.  The fermion number
$F=\hat{\pi}_1\hat{\psi}_1+\hat{\pi}_2\hat{\psi}_2$ is a conserved
quantity of this Hamiltonian.  After the time evolution from
$t=-\infty$ only the ground state with the lowest energy eigenvalue
remains, so that the generalized weight is given by
$P(z,\bar{z})\rmd z\rmd \overline{z}=\Psi(z,\bar{z})\wedge\rme^{S(z)}\rmd z$, where
$\Psi(z,\bar{z})$ is the ground state wave-function and converges to
$\delta(\calJ_\sigma)$ in the limit $\ve_i\to +0$.  
Note that the weight factor $\exp S(z)$ is necessary in this formula, since the wave function designates only the integration cycle $\mathcal{J}_{\sigma}$. 
We can further simplify this Hamilton problem by choosing $\ve=\ve_1=\ve_2$.  
Performing the conjugate transformation $\Psi=\rme^{-\Re S/\ve}\Psi'$, the first derivative terms are eliminated as
\begin{equation}
  H_{\rm eff}' = \sum_i \biggl[ \frac{\ve}{2}\hat{p}_i^2
  + \frac{1}{2\ve}\biggl(\frac{\partial \Re S}{\partial x_i}
  \biggr)^2 \biggr] - \sum_{i,j}\frac{1}{2}\frac{\partial^2 \Re S}
  {\partial x_i \partial x_j}\bigl[ \hat{\pi}_i,\hat{\psi}_j\bigr] \;.
\label{eq:Heff_susy}
\end{equation}
This describes supersymmetric quantum mechanics with the
superpotential $\Re S$~\cite{Witten:2010zr}.

Before applying this method to an interacting model, let us convince
ourselves of perturbative correctness.  For that purpose, we consider the simple Gaussian case:
\begin{equation}
  S_0(x) = -\frac{\omega}{2} x^2 \;,
\label{eq:S_free}
\end{equation}
where $x\in\mathbb{R}$ is a one-component variable and
$\omega\in\mathbb{C}$.  One can regard this Gaussian integral as
  an elementary building-block of perturbative quantum field theory;  in a
  non-interacting theory of the scalar field $\phi$, the action is
  decomposed into $-\frac{1}{2}[\Gamma-\rmi(k^2-m^2)]\phi(-k)\phi(k)$
  for each Fourier mode.
In this case we can immediately find that the bosonic wave-function
should be the ground state of harmonic oscillator with the ground
state energy $|\omega|$.  The fermionic ground state energy
$-|\omega|$ cancels the bosonic energy thanks to supersymmetry. 
As a result, unoccupied fermions point the direction of the
Lefschetz thimble; i.e., the supersymmetric vacuum belongs to the
$F=1$ sector.  
Let us see this in the simplest example, $\omega\in\mathbb{R}$ and $\omega>0$. 
The fermionic part of the Hamiltonian is $\frac{\omega}{2}([\hat{\pi}_1,\hat{\psi}_1]-[\hat{\pi}_2,\hat{\psi}_2])$,
 and thus the $1$ and $2$ fermions are unoccupied and occupied, respectively. 
This leads to the fermionic ground state energy
  $-\omega$, which cancels the bosonic ground state energy $\omega$. 
Therefore, the unoccupied fermion is tangent to the Lefschetz thimble $\mathcal{J}=\mathbb{R}$. 

The final result of $P_0(z,\bar{z})$ for
$\omega\in\mathbb{C}$ together with
$\rme^{S}\cdot\rme^{-\frac{1}{\varepsilon}\Re S}$ is
\begin{equation}
 P_0(z,\bar{z}) = \calN\exp\biggl[
  -\frac{|\omega|}{2\varepsilon}z\bar{z} + \frac{1}{4\varepsilon}
  (\omega z^2+\bar{\omega}\bar{z}^2)-\frac{\omega}{2}z^2 \biggr]\;,
\label{eq:P_lt}
\end{equation}
which reproduces the original integral~\eqref{eq:original} with the
  action~\eqref{eq:S_free}.  
  To see this for a polynomial $\calO(x)$
  it is sufficient to require $\langle z^2\rangle=1/\omega$, which is
  nothing but the free propagator and can be explicitly confirmed with
  Eq.~\eqref{eq:P_lt}.
In order for the exponentially fast convergence of $P_0$, the
parameter $\ve$ needs to be $0\le \ve<2$.  We here emphasize that
  the theory is equivalent to the original~\eqref{eq:original} for
  $0\le\forall\ve<2$ and the conventional Lefschetz thimble method is
  retrieved in the $\ve\to0$ limit.
Actually, in this limit of $\ve\to 0$, only a path of
$2|\omega|z\bar{z}-\omega z^2-\bar{\omega}\bar{z}^2 \;(\equiv
\{2{\rm Im}(\sqrt{\omega}z)\}^2)=0$ contributes,
which is nothing but a condition to guarantee $\Im S_0(z)=0$ on the
Lefschetz thimble.  At finite $\ve$, this restriction is smeared and
$P_0$ may have a distribution around $\calJ$ with a width of order
  of $\ve$ where a complex phase arises in general. 
The non-positivity of $P_0$ is a big difference of this Lefschetz-thimble approach from the complex Langevin method, in which the distribution must be semi-positive definite. 

In the model with quartic interaction,
$S=-{\omega\over 2}z^2-{\lambda\over 4}z^4$, the vacuum structure is
drastically different from that of the Gaussian case~\eqref{eq:P_lt}.
With $\lambda\neq 0$, there are three classical saddle points and the
Morse index for each saddle point is $1$.   The Witten index is
$\mathrm{tr}(-1)^F=-3$, and thus there are three supersymmetric vacua
in this interacting model for any
$\ve$~\cite{Witten:1982im,Behtash:2015kva}.  In the path integral
expression~\eqref{eq:path_integral_expression_thimble}, we can
distinguish these three vacua by specifying the boundary condition at
$t=-\infty$, as we will discuss below in detail.

\paragraph{3.\ Lefschetz thimbles at $\ve\to 0$.}

We define our zero-dimensional model with the quartic interaction by
\begin{equation}
  S(x) = -{\omega\over 2}x^2-{\lambda\over 4}x^4,
\label{eq:model}
\end{equation}
and hereafter, we will specifically choose the model parameters as
\begin{equation}
 \omega = 1-\rmi=\sqrt{2}{\rme^{-\rmi\pi/4}},\qquad
 \lambda = 1.5\rmi\;.
\label{eq:param}
\end{equation}
This choice has been motivated by the application to real-time
problems.  As we already mentioned, $\Re\omega$ corresponds to the
width $\Gamma$ or $\epsilon$ in the $\rmi\epsilon$ prescription, and
$\Im\omega$ corresponds to $-(k^2-m^2)$.  Therefore, the
parameters~\eqref{eq:param} represent a situation at $k^2=m^2+\Gamma$
in quantum field theory.

\begin{figure}\centering
 \includegraphics[width=0.4\columnwidth]{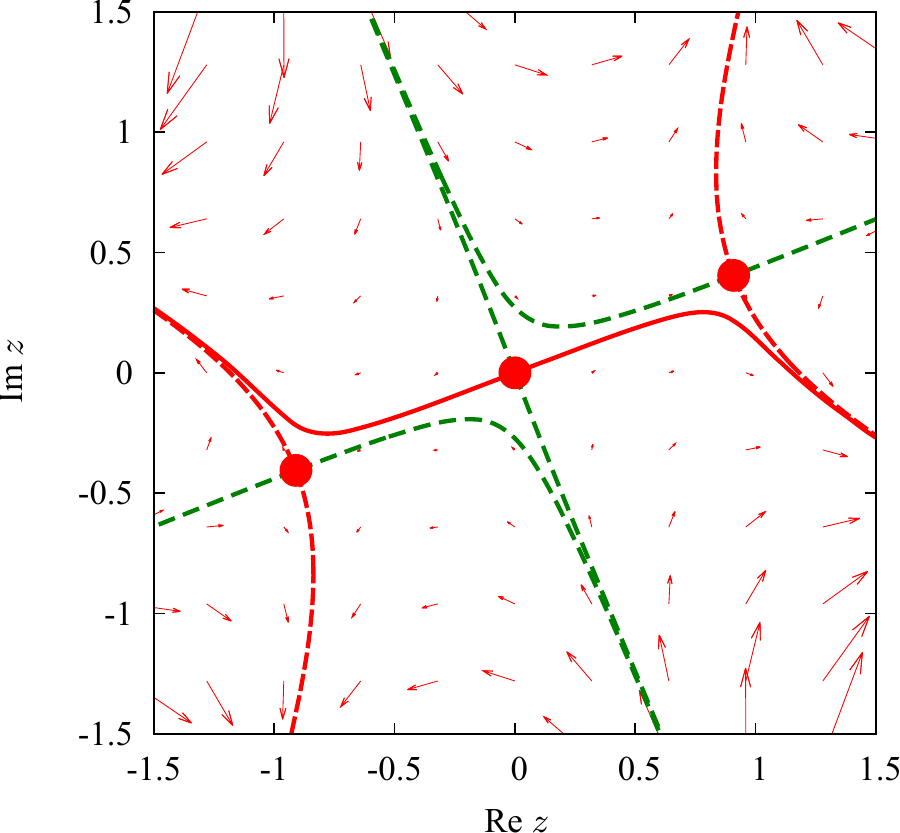} \hspace{1em}
 \includegraphics[width=0.4\columnwidth]{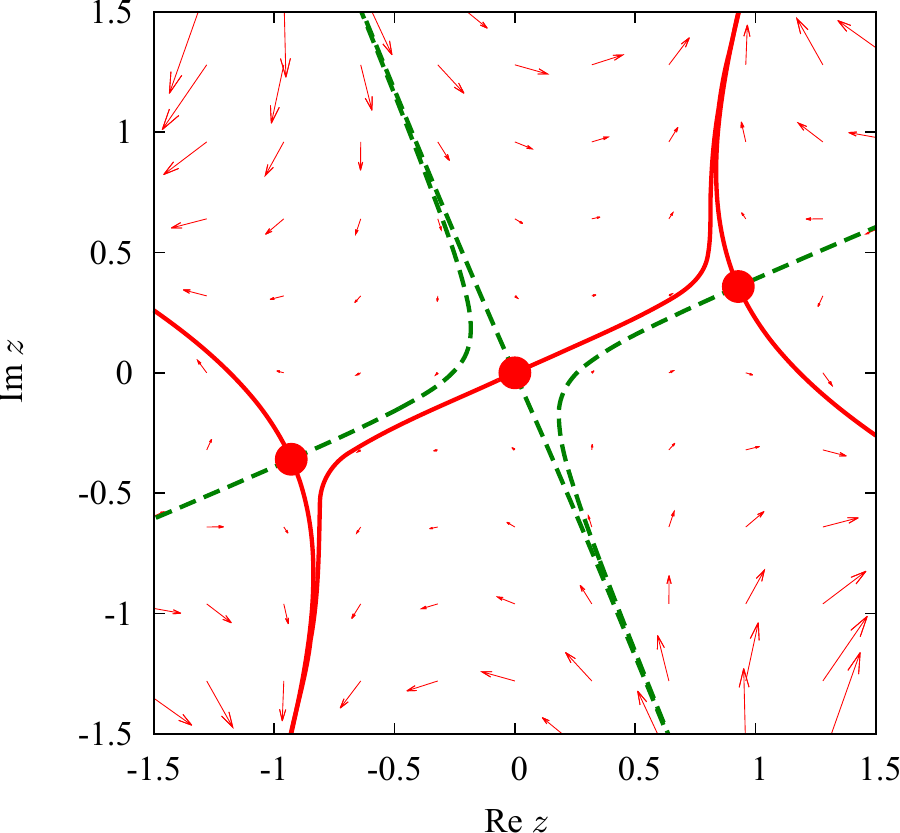}
 \caption{Changes of the Lefschetz thimbles for $\omega=1-0.9\rmi$
   (left) and $\omega=1-1.1\rmi$ (right) with $\lambda=1.5\rmi$ fixed.
   (Left) One of three thimbles as shown by the solid line contributes
   to the integral and two are to be dropped when $\omega=1-0.9\rmi$.
   (Right) All three thimbles contribute to the integral when
   $\omega=1-1.1\rmi$.}
 \label{fig:thimble}
\end{figure}

When $\ve$ is small enough, the Hamilton dynamics should become
equivalent to the conventional formulation of the integral on the
Lefschetz thimble, which means that $P(z,\bar{z})$ should have a peak
along the Lefschetz thimble only.  This makes the analytic treatment
much accessible since we do not have to solve the quantum mechanical
problem in this zero-dimensional toy model.  To identify the Lefschetz
thimble, we should integrate the flow equation in
Eq.~\eqref{eq:thimble} and find the upward and the downward paths.  We
show the numerical results in Fig.~\ref{fig:thimble} for parameters
slightly changed from Eq.~\eqref{eq:param}.

For our theory~\eqref{eq:model} three saddle points are located at
$z_0=0$, $z_\pm=\pm\sqrt{-\omega/\lambda}$.  The latter is, for our
choice of parameters~\eqref{eq:param},
$z_\pm=\pm(0.897+0.372\rmi)$, and $S(z_{\pm})=-1/3$.
We can see that $\omega=1-\rmi$ is a critical value at which the
destinations of the downward flows from the saddle points change
drastically, that is known as the Stokes phenomenon, as is clear from
two panels for $\omega=1-0.9\rmi$ (left) and $\omega=1-1.1\rmi$
(right) in Fig.~\ref{fig:thimble}.  In general we can show that the
Stokes phenomenon occurs at $\Im(\omega^2/\lambda)=0$, and when
$\lambda$ is pure imaginary, this condition gives
${\rm arg}\,(\omega)=-\pi/4$.  Importantly, not only the downward
flows but also the upward flows change and the intersection number of
the original and the upward paths changes
accordingly~\cite{Witten:2010cx,Witten:2010zr}.  
In the case with $\omega=1-0.9\rmi$ only one upward path from $z_0$
crosses the real axis as seen in the left of Fig.~\ref{fig:thimble},
and so the Lefschetz thimble going through $z_0$ contributes to the
integral.  In the case with $\omega=1-1.1\rmi$, on the other hand,
three upward paths from $z_0$ and $z_\pm$ all cross the real axis, and
all three Lefschetz thimbles contribute to the integral.

Keeping the potential application to the real-time physics in
  mind, we need deeper understanding on the Stokes phenomenon.  In
  fact, it occurs at $k^2=m^2+\Gamma$ and so the thimble structure
  may fluctuate depending on the frequency $k^0$, while in Euclidean
  theory it never happens because $k^2-m^2$ is always negative (except
  for an unstable potential with $m^2<0$).  It should be an
  interesting future problem to clarify the treatment of the Stokes
  phenomenon in the real-time systems \cite{Tanizaki:2014xba}.

For the same model with a different set of parameters, the Stokes
phenomenon has been discussed in
Refs.~\cite{Aarts:2013fpa,Aarts:2014nxa} and the probability
distribution $P(z,\bar{z})$ in the complex Langevin method has been
numerically computed.  The important insight obtained there is that
$P(z,\bar{z})$ in the complex Langevin method looks localized but has
a power decay at large $|z|$, which causes a convergence problem.
Hence, it would be an intriguing question how $P(z,\bar{z})$ in the
modified Lefschetz thimble method for $\ve\neq0$ should behave
especially at large $|z|$.

\paragraph{4.\ Wave-functions at finite $\ve$.}

As an application of the modified Lefschetz-thimble method with a
regulator $\ve$, let us compute the wave-function $\Psi'$ for the
$\ve=\ve_1=\ve_2$ case, from which $P(z,\bar{z})$ is to be constructed
immediately.  We can readily find the eigenstate in terms of $\psi_i$
for the last term in the Hamiltonian~\eqref{eq:Heff_susy}.  By
restricting ourselves to the $F=1$ sector, we can define the effective
potential in a form of $2\times 2$ matrix-valued function that amounts
to
\begin{equation}
  V_{\rm eff} = \frac{1}{2\ve}\biggl[\biggl(\frac{\partial \Re S}
    {\partial x_1}\biggr)^2 + \biggl(\frac{\partial \Re S}
    {\partial x_2}\biggr)^2 \biggr]
  - \biggl(
  \begin{array}{cc}
    {\partial^2 \Re S/\partial x_1^2}
    & {\partial^2 \Re S/\partial x_1\partial x_2}\\
    {\partial^2 \Re S/\partial x_1\partial x_2}
    &  -{\partial^2 \Re S/\partial x_1^2}
 \end{array}\biggr)\;,
\label{eq:Veff}
\end{equation}
and then the Hamiltonian is
\begin{equation}
  H_{\rm eff}' = -\frac{\ve}{2}\biggl(\frac{\partial^2}{\partial x_1^2}
  +\frac{\partial^2}{\partial x_2^2}\biggr) + V_{\rm eff} \;.
\label{eq:HeffVeff}
\end{equation}
Let us solve the ground state of the above $H_{\rm eff}'$ at finite
$\ve$, and we specifically adopt $\ve=1$ unless stated explicitly.  

When we solve the Hamilton dynamics, we should set initial conditions
to select proper Lefschetz thimbles out.  
We choose the semiclassical ground state in the limit $\varepsilon\to+0$ as our initial condition. 
Since $\Psi'=\mathrm{e}^{\mathrm{Re}S/\varepsilon} \Psi$ and $\Psi^{(z_{\sigma})}\to \delta(\mathcal{J}_{\sigma})$ in $\varepsilon\to+0$, $\Psi'^{(z_{\sigma})}\sim \mathrm{e}^{\mathrm{Re}S/\varepsilon}\delta(\mathcal{J}_{\sigma})$ for small $\varepsilon$. 
Let us consider the Lefschetz thimble around $z_0$ for instance, then
its tangential direction at $z_0$ is given by $x_2=\tan{(\pi/8)}x_1$
as seen also from Fig.~\ref{fig:thimble}. 
The initial wave-function is thus proportional to
\begin{equation}
  \mathrm{e}^{\mathrm{Re}S(z)/\varepsilon}\delta(-\sin(\pi/8)x_1+\cos(\pi/8)x_2) \cdot
  (-\sin(\pi/8)\rmd x_1+\cos(\pi/8)\rmd x_2) \;. 
\label{Eq:WKB_wavefunction_origin}
\end{equation}
The bosonic part is well localized at $z_0$, which justifies the following choice of the initial wave function, $\Psi'^{(z_\sigma)}(t=-\infty)=\delta(z-z_0) (-\sin(\pi/8)\rmd x_1+\cos(\pi/8)\rmd x_2)$. 
Similarly, we can fix the initial wave function for $z_{\pm}$ as 
$\Psi'^{(z_{\pm})} =\delta(z-z_{\pm}) (\cos(\pi/8)\rmd x_1+\sin(\pi/8)\rmd{x_2})$.

\begin{figure}
\begin{minipage}{.5\textwidth}
\subfloat[$\Psi'^{(z_0)}_1(x_1,x_2)$]{
 \includegraphics[width=0.9\columnwidth]{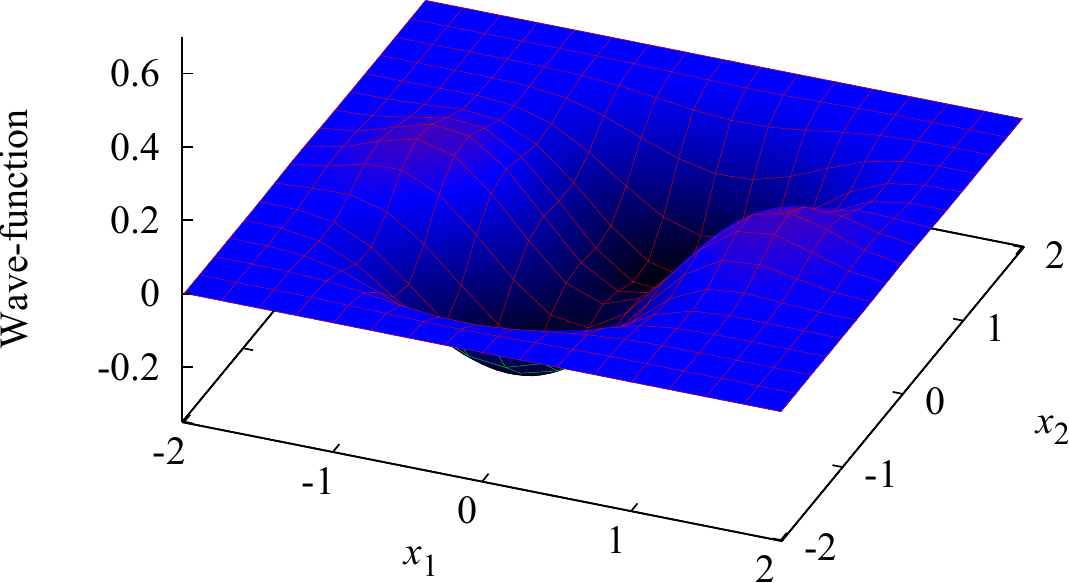}} 
\end{minipage}
\begin{minipage}{.5\textwidth}
\subfloat[$\Psi'^{(z_0)}_2(x_1,x_2)$]{
 \includegraphics[width=0.9\columnwidth]{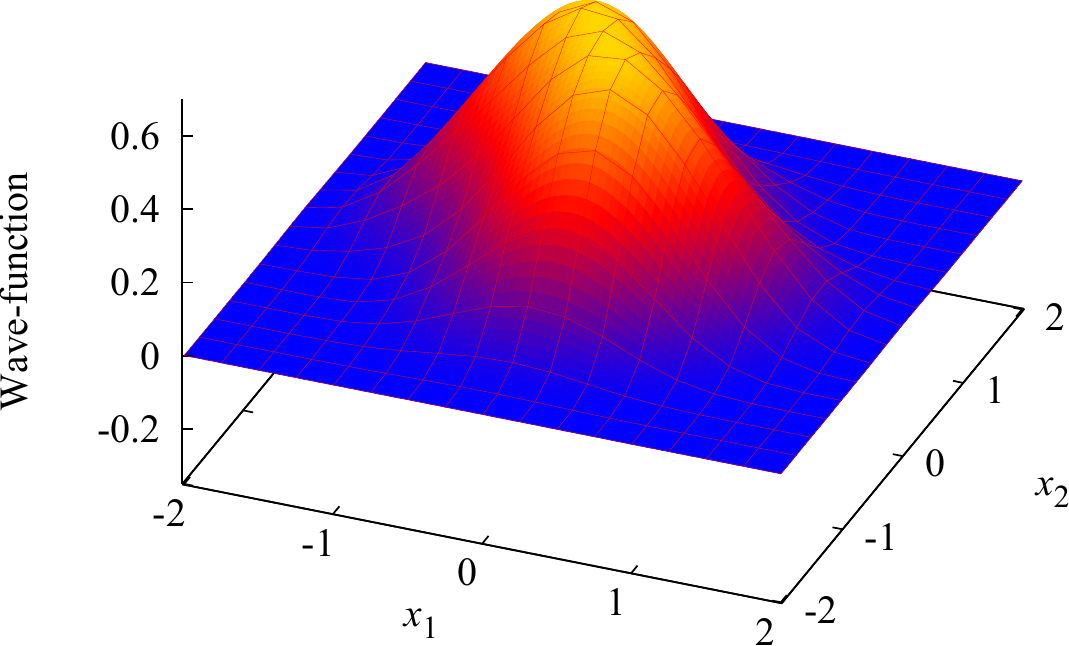}}
\end{minipage}
\begin{minipage}{.5\textwidth}
\subfloat[$\Psi'^{(z_+)}_1(x_1,x_2)$]{
 \includegraphics[width=0.9\columnwidth]{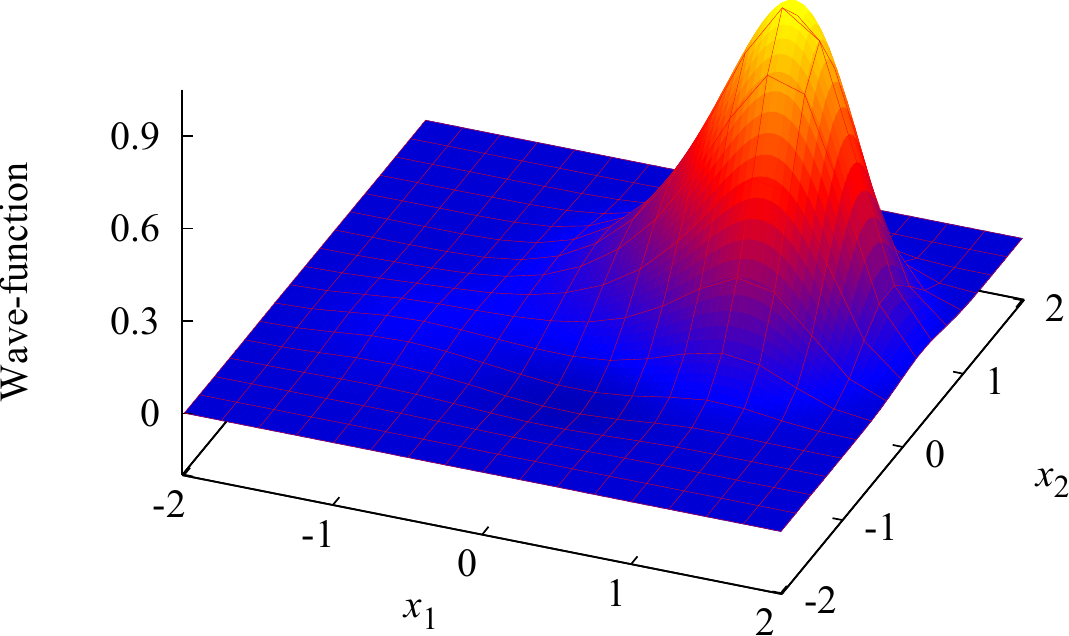}}
\end{minipage}
\begin{minipage}{.5\textwidth}
\subfloat[$\Psi'^{(z_+)}_2(x_1,x_2)$]{
 \includegraphics[width=0.9\columnwidth]{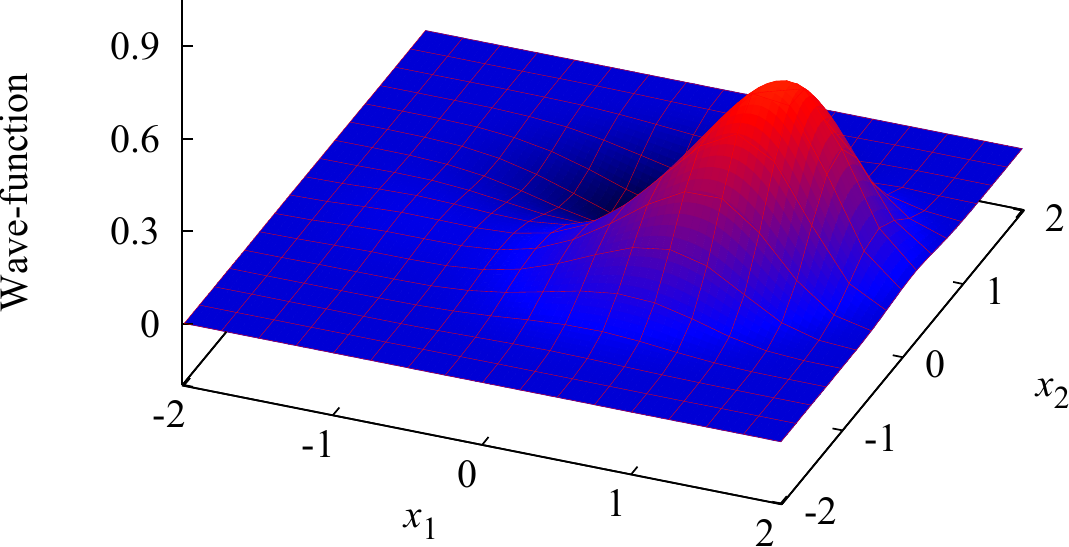}}
\end{minipage}
 \caption{SUSY ground state $\Psi'^{(z_{\sigma})}$ corresponding to
   the saddle points
   $z_0=0$ and $z_+=\sqrt{-\omega/\lambda}$ with $\varepsilon=1$,
   $\omega=1-\rmi$, and $\lambda=10^{-3}+1.5\rmi$.  Denoting
   $\Psi'^{(z_{\sigma})}=\Psi'^{(z_{\sigma})}_1\rmd x_1
   + \Psi'^{(z_{\sigma})}_2\rmd x_2$,  (a) and (b) represent
   $\Psi'^{(z_0)}$, while (c) and (d) represent $\Psi'^{(z_+)}$.}
  \label{fig:thimble_wavefunction}
\end{figure}

For the numerical procedure we smeared the delta function in the initial wave function by
  a Gaussian as $\delta(z-z_\sigma) \to
  \exp\{-20[(x_1-x_{1\sigma})^2+(x_2-x_{2\sigma})^2]\}$.  Then we discretized
  $x_1$ and $x_2$ from $-2.5$ to $+2.5$ with $\rmd x=5\times 10^{-2}$.
  We then numerically integrate
  ${\rmd \over \rmd t}{\Psi'^{(z_{\sigma})}}=-H_{\rm eff}'\Psi'^{(z_{\sigma})}$ using
  the Euler method with $\rmd t=10^{-4}$ until the
  wave-function converges.  The convergence is fast and stable and the
  wave-function hardly changes after $t= 1\sim 2$.  We also mention
  that we utilized the Crank-Nicolson algorithm to improve the
  numerical stability when we compute the Laplacian.
  
With this prescription, we find three {independent} ground state
wave-functions $\Psi'^{(z_{\sigma})}=\Psi'^{(z_{\sigma})}_1\rmd x_1+\Psi'^{(z_{\sigma})}_2\rmd x_2$, which is shown in
Fig.~\ref{fig:thimble_wavefunction}.  The figures (a,b) in
Fig.~\ref{fig:thimble_wavefunction} show two components of
$\Psi'^{(z_0)}$ and the figures (c,d) show those of $\Psi'^{(z_+)}$.
Since our system is symmetric under the reflection $z\mapsto -z$, we
can easily read $\Psi'^{(z_-)}$ from the figures of $\Psi'^{(z_+)}$.
Remarkably, these ground-state wave functions are linearly independent from one another, which means that supersymmetry is unbroken.
We have also verified that the Dyson--Schwinger equation, 
\begin{equation}
\lambda \langle z^4\rangle + \omega \langle z^2\rangle=1, 
\label{eq:Dyson_Schwinger_equation}
\end{equation}
is satisfied within $1\%$ accuracy. This clearly shows that the supersymmetric quantum mechanics provides a suitable framework to compute Lefschetz thimbles. 

Let us comment on the convergence of the wave-functions in the
  present modified Lefschetz thimble method.  In the effective
  potential~\eqref{eq:Veff}, the first term gives the dominant binding potential in our toy
  model for large $|z|$ because the first term is a polynomial up to
  the sixth order, while the second term is up to the quadratic order.
Therefore, the convergence in the present numerical
  approach is quite improved and there is no problem of the
  power-decay unlike the complex Langevin method.  One might think
  that the remaining $\rme^{\rmi\Im S}$ is still oscillating, but this
  is not a problem practically.  We have numerical verified that the
  effect of $\rme^{\rmi\Im S}$ is very small in the region where the
  profile of $\Psi'^{(z_\sigma)}$ is localized.

\begin{figure}[t]
\begin{minipage}{.5\textwidth}
\subfloat[$\tilde{\Psi}'^{(z_0)}_1(x_1,x_2)$]{
 \includegraphics[width=0.9\columnwidth]{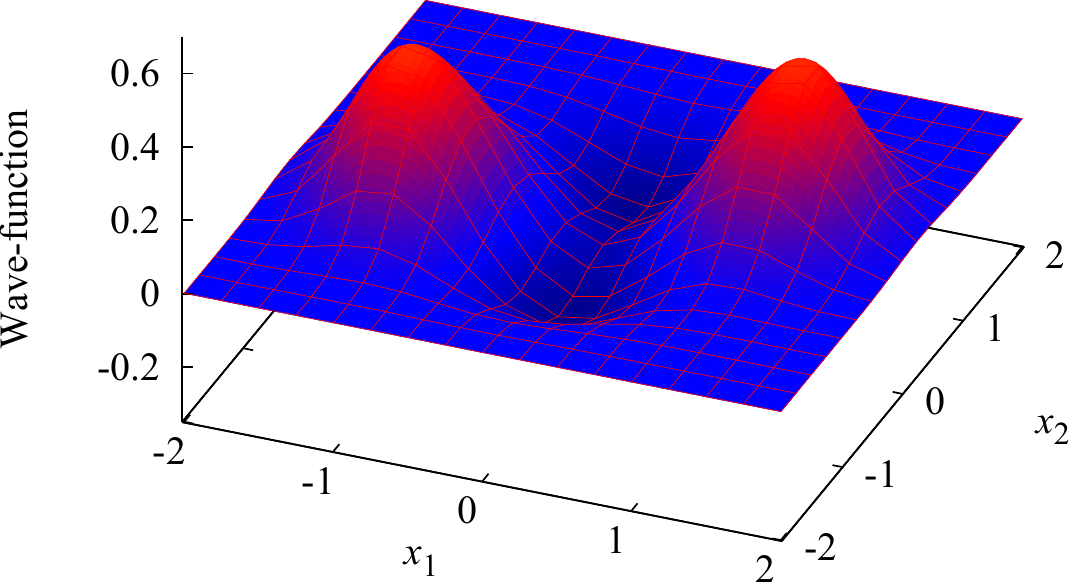}} 
\end{minipage}
\begin{minipage}{.5\textwidth}
\subfloat[$\tilde{\Psi}'^{(z_0)}_2(x_1,x_2)$]{
 \includegraphics[width=0.9\columnwidth]{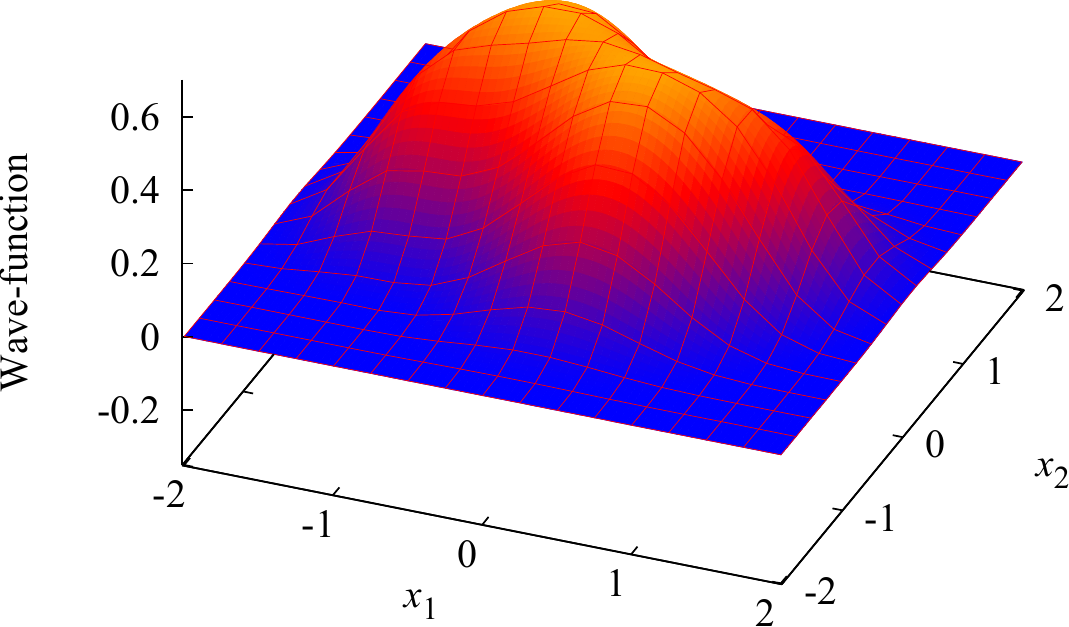}}
\end{minipage}
 \caption{SUSY ground state $\tilde{\Psi}'^{(z_{0})}$, which is
   obtained starting with the saddle point $z_0$ but with the initial
   relative weight equal to both components.  
} \label{fig:thimble_wavefunction_equalweight}
\end{figure}

We also checked the robustness of the numerical results against
  small variations of the center position and the smearing width in
  initial conditions.  If the smearing width of the initial wave function is changed, the overall
  normalization would be changed naturally, but once we normalize the
  wave-functions in the same manner (in
  Fig.~\ref{fig:thimble_wavefunction} we normalized them as
  $\int\rmd x_1\,\rmd x_2[(\Psi'^{(z_\sigma)}_1)^2+(\Psi'^{(z_\sigma)}_2)^2]=1$),
  then we eventually get the same result.  Such insensitivity implies
  that each wave-function is well localized and the overlap at the
  saddle point is small.
However, the overlap at the saddle point is not completely zero.  To
see this effect, let us skew the relative weight of the initial
condition so that the initial wave-functions are not orthogonal.
In Fig.~\ref{fig:thimble_wavefunction_equalweight}, we show
$\tilde{\Psi}'^{(z_0)}_1$ and $\tilde{\Psi}'^{(z_0)}_2$ starting with
  the $\text{same}$ relative weights, namely,
  $\tilde{\Psi}'^{(z_0)}(-\infty)=\delta(z-z_0)(\rmd x_1+\rmd x_2)$ at the
  initial time, for the demonstration purpose.
The result $\tilde{\Psi}'^{(z_0)}$ in
Fig.~\ref{fig:thimble_wavefunction_equalweight} is given a natural
  interpretation as a superposition of three ground states shown in
  Fig.~\ref{fig:thimble_wavefunction}. 
Because of this overlap among wave functions, our prescription for the initial wave function needs further refinement in order to extract one Lefschetz thimble at $\varepsilon=1$. 

\begin{figure}
\begin{minipage}{.5\textwidth}
\subfloat[$\mathrm{Re}P$ at $\omega=1-\rmi$, $\varepsilon=0.2$]{
 \includegraphics[width=0.9\columnwidth]{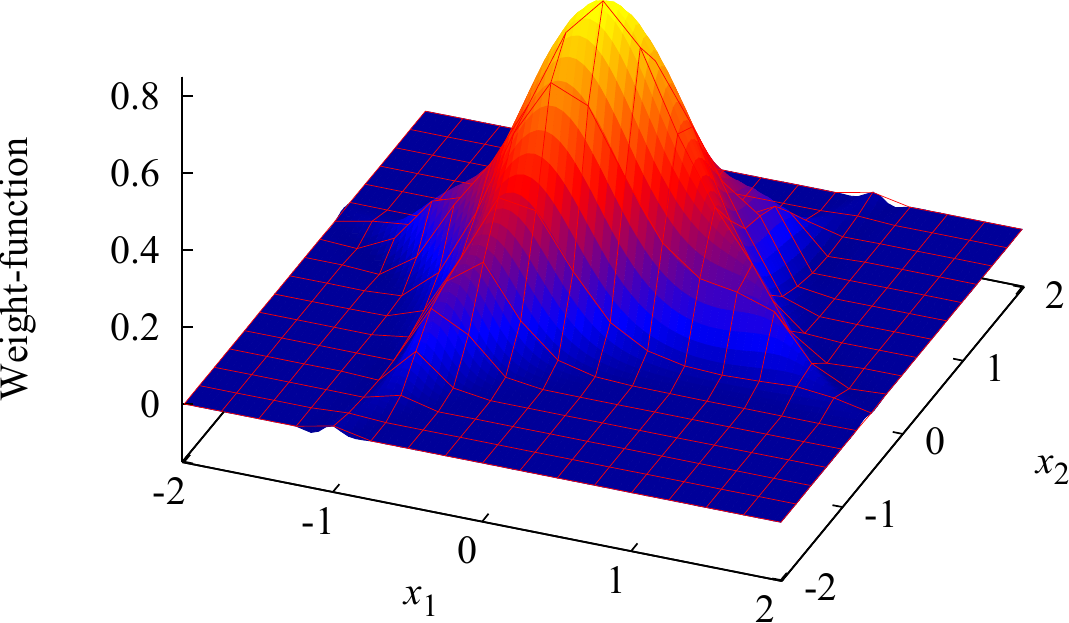}} 
\end{minipage}
\begin{minipage}{.5\textwidth}
\subfloat[$\mathrm{Im}P$ at $\omega=1-\rmi$, $\varepsilon=0.2$]{
 \includegraphics[width=0.9\columnwidth]{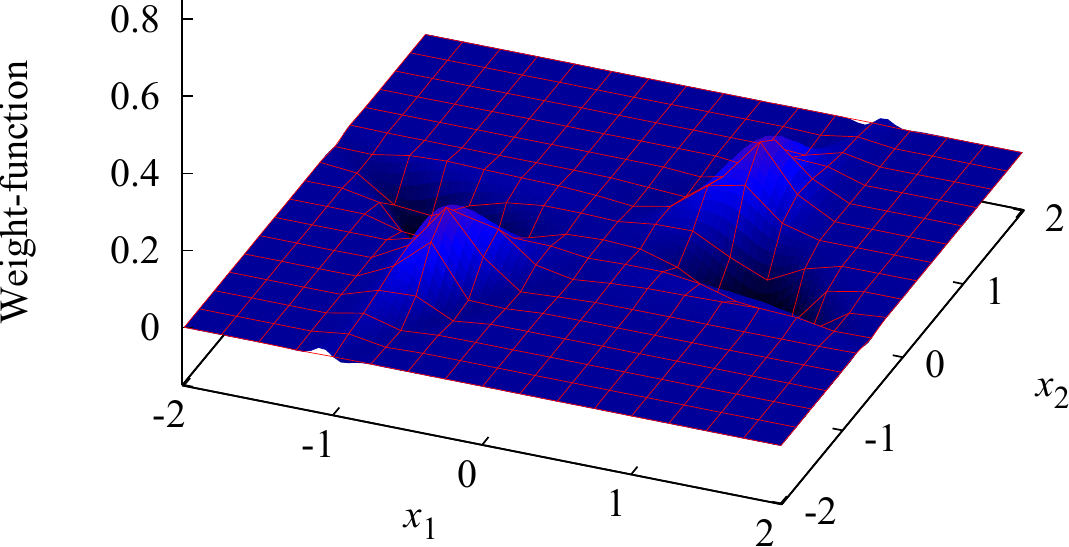}}
\end{minipage}
\begin{minipage}{.5\textwidth}
\subfloat[$\mathrm{Re}P$ at $\omega=1+\rmi$, $\varepsilon=0.2$]{
 \includegraphics[width=0.9\columnwidth]{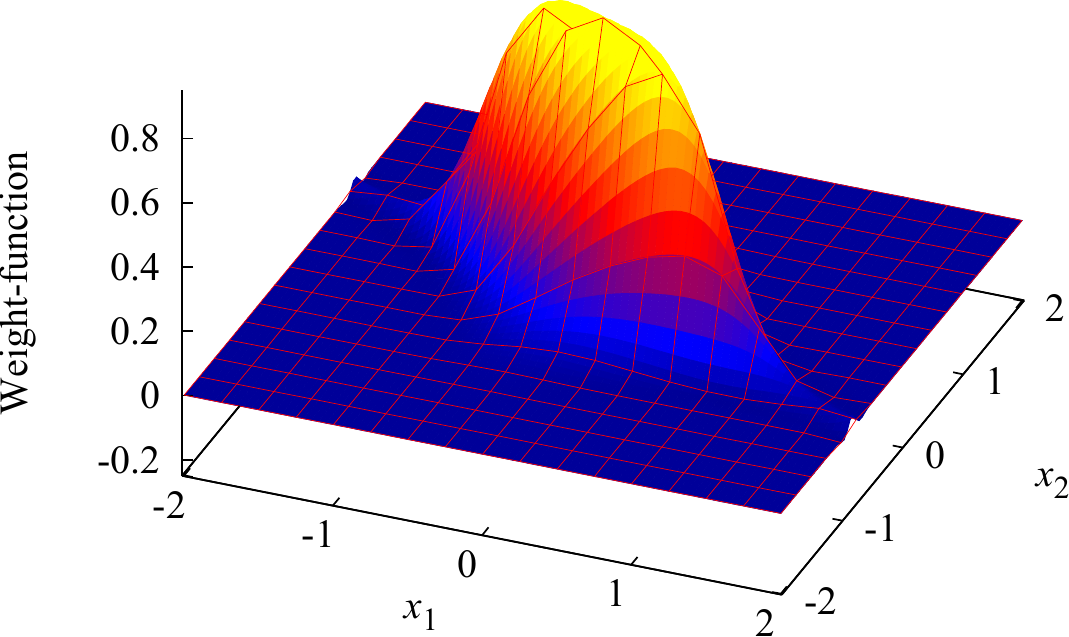}}
\end{minipage}
\begin{minipage}{.5\textwidth}
\subfloat[$\mathrm{Im}P$ at $\omega=1+\rmi$, $\varepsilon=0.2$]{
 \includegraphics[width=0.9\columnwidth]{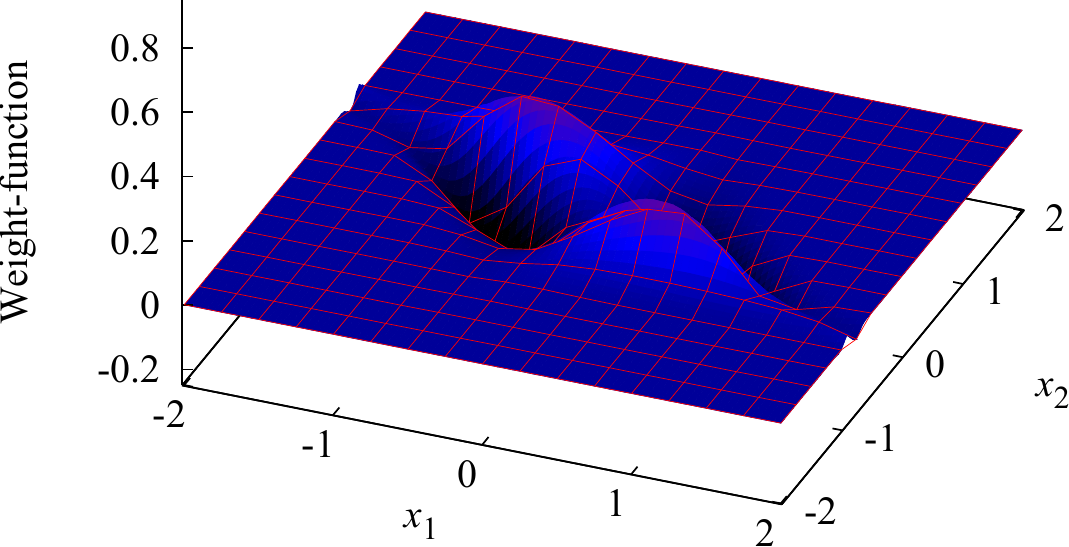}}
\end{minipage}
 \caption{Weight functions $P(x_1,x_2)$ corresponding to
   the saddle point $z_0$ with $\varepsilon=0.2$. (a)~and~(b) show the result for $\omega=1-\rmi$, and $\lambda=10^{-3}+1.5\rmi$. (c)~and~(d) show the result for $\omega=1+\rmi$, and $\lambda=10^{-3}+1.5\rmi$. }
  \label{fig:thimble_weightfunction}
\end{figure}

In order to see the importance of such refinements from another viewpoint, we checked behaviors of distributions $P$ for small $\varepsilon$. 
Let us set $\varepsilon=0.2$ with $\omega=1-\rmi$ and $\lambda=1.5\rmi$, then we obtain Figs.~\ref{fig:thimble_weightfunction} (a) and (b). 
Here, the normalization is given by $\int \rmd^2 x\,P=1$. 
Although it is localized around the saddle point $z_0=0$, its shape is still far from a one-dimensional line shown in Fig.~\ref{fig:thimble}. 
In order for comparison, we show the result also for $\varepsilon=0.2$ with $\omega=1+\rmi$ and $\lambda=1.5\rmi$ in Figs.~\ref{fig:thimble_weightfunction} (c) and (d). 
For this case, $P$ is well localized to a one-dimensional line, which is nothing but the Lefschetz thimble $\mathcal{J}_0$ at this parameter. 
We numerically observed that the weight function $P$ is well localized around a Lefschetz thimble if Stokes phenomena do not happen and $\varepsilon$ is sufficiently small. 
Expectation values of operators, such as $\langle z^2\rangle$, also give correct numbers within the numerical accuracy with such parameters. 
However, as $\varepsilon$ becomes larger, wave functions spread as two-dimensional distributions, and expectation values of operators do not necessarily give correct values. 
Since the Dyson--Schwinger equation (\ref{eq:Dyson_Schwinger_equation}) is satisfied, this problem must come from superpositions of the wave function with other ground states. 

It is an important future study to clarify dependence on this initial condition more systematically, in order to take an appropriate linear combination of these wave-functions giving each Lefschetz thimble. 
This would be a key step to study how the Stokes phenomenon is realized at $\varepsilon=1$. 
It also opens a new possibility to take into account multiple Lefschetz thimbles, 
since ground-state wave functions can be superposed by changing the initial condition.

\paragraph{5.\ Discussions and Conclusions.}

In this Letter, the supersymmetric reformulation of the Lefschetz
thimble integration was studied and its practical computation was
discussed.  The Lefschetz thimbles are now regarded as the ground
sates of a supersymmetric Hamiltonian.  
Our computational scheme for the Hamiltonian system is essentially the same with that for the Fokker-Planck equation in the complex Langevin method.  
Those supersymmetric wave-functions are numerically computed for a zero-dimensional toy
model with the classical action $S=S_0+S_{\rm int}$ where
$S_0=-\frac{\omega}{2}x^2$ and $S_{\rm int}=-\frac{\lambda}{4}x^4$.
Since the Morse indices of all the saddle points are the same, the
number of saddle points must be the same as that of linearly
independent ground states.

For the Gaussian model with $S_0$ only, the above-mentioned situation
is clearly true, which is explicitly checked by computing the
wave-function analytically and constructing one supersymmetric ground
sate.  From this almost trivial example, one can learn an important
lesson about the two-dimensional smooth distribution $P_0(z,\bar{z})$.
In the ``semi-classical'' limit of $\ve\to0$, we recover a delta
functional support along the Lefschetz thimble in the original
formulation.  For non-zero $\ve$, the phase oscillation arises away
from the Lefschetz thimble, and the formulation nevertheless
reproduces the correct expectation value.

For the interacting model with $S_{\rm int}$, we performed numerical
computations for the supersymmetric quantum mechanics, and confirmed
the existence of three linearly independent ground states by
restricting ourselves to the $F=1$ sector.  Since the wave-function in
the $F=1$ sector consists of two components, we must set them in the
proper initial conditions.  We started from a localized wave-function
in the vicinity of each saddle point, and our numerical computation
shows a remarkable stability under small modifications on the initial
conditions.  This reflects the fact that all the saddle points are
  attractive unlike the complex Langevin method.
At the same time we also found substantial dependence on the initial
relative weight of these two components.  This clearly indicates that
we need a careful refinement of the initial condition to use the
modified Lefschetz thimble method for numerical simulations, but it
also opens a new possibility for some convenient scheme to take
into account multiple Lefschetz thimbles.

In the case of the Fokker-Planck system, the ground state is
often uniquely determined and the initial condition dependence does
not appear as in the case in ordinary quantum mechanics without any
special symmetry. 
The Fokker-Planck operator of the complex Langevin method can be written as a
functional integral in a similar manner, and it also shows the same
type of the BRST symmetry. 
There are, however, several important
differences in these two formalisms: 
First of all, the diffusion term in the complex Langevin equation does not give the gradient flow because the sign is different. 
Because of this difference,
the BRST invariance cannot be promoted to the supersymmetric system
satisfying $2H=\{Q,\bar{Q}\}$ with supercharges $Q,\,\bar{Q}$. Moreover, the
fermionic sector should be restricted to $F=2$ instead of $F=1$.
Therefore, it is not straightforward to relate these two formalisms yet. 
If one could establish a
firm relation between two methods, it would be a great progress and
the present modified Lefschetz thimble could provide us with an
opportunity for a fully complementary approach to the sign problem
together with the complex Langevin method.

\ack
The authors thank Yuya~Abe, So~Matsuura, Jun~Nishimura, and
Jan~Pawlowski for useful discussions.  K.~F.\ was supported by JSPS
KAKENHI Grant No.\ 15H03652 and 15K13479.
Y.~T.\ was supported by Grants-in-Aid for JSPS fellows (No.25-6615). 
This work was partially supported by  the RIKEN iTHES project, and
also by the Program for Leading Graduate Schools, MEXT, Japan.


\bibliographystyle{ptephy}
\bibliography{lefschetz}
\end{document}